\documentclass[floatfix,
reprint, amsmath, amssymb,
aps,pra]{revtex4-2}

\usepackage{graphicx}
\usepackage{dcolumn}
\usepackage{bm}

\begin{document}

\title{Photon-count fluctuations exhibit inverse-square baseband spectral behavior\\ that extends to $\bm{< 1\;\mu}$Hz}

\author{Nishant~Mohan}
 \altaffiliation[Also with ]{Boston University, Department of Biomedical Engineering, Boston, Massachusetts 02215, USA; Massachusetts General Hospital, Wellman Center for Photomedicine, Boston, Massachusetts 02114, USA; and Wasatch Photonics, Morrisville, North Carolina 27560, USA}
\author{Steven~B.~Lowen}
 \altaffiliation[Also with ]{Boston University, Department of Electrical \& Computer Engineering, Boston, Massachusetts 02215, USA; and Harvard Medical School, Boston, Massachusetts 02115, USA}
\author{Malvin~Carl~Teich}
 \altaffiliation[Also with ]{Boston University, Departments of Electrical \& Computer Engineering, Biomedical Engineering, and Physics, Boston, Massachusetts 02215, USA; and Columbia University, Department of Electrical Engineering, New York, New York 10027, USA; teich@bu.edu}
\affiliation{%
Boston University, Photonics Center, Boston, Massachusetts 02215, USA
}%

\date{\today}%

\begin{abstract} A broad variety of light sources exhibit photon-count fluctuations that display inverse-square spectral behavior at extremely low frequencies. These sources include light-emitting diodes, superluminescent diodes, laser diodes, incandescent sources, and betaluminescent sources. In a series of experiments carried out over an 18-month period, the photon-count fluctuations for these sources were found to exhibit a $1/f^2$ spectral signature over the frequency range $1.0 \times 10^{-6} \le f \le 5.0 \times 10^{-4}$~Hz, corresponding to $33$~min $ \le T_f \le 11.6$~d, where $T_f \equiv 1/f$. The lower time limit is established by the photodetector noise floor while the upper time limit is determined  by the duration of the individual experiments. Scalograms computed from our data are consistent with the periodograms. The universal character of this inverse-square baseband spectral behavior stands in sharp contrast to the optical spectra of these sources, which differ markedly. Time traces of the photon-count fluctuations are examined to shed light on the origin of this enigmatic spectral behavior.
\end{abstract}

\maketitle

\section{Introduction}

Photon-count fluctuations, an intrinsic feature of light, play an important role in many physical and biological processes. Their presence impacts the design of experiments in areas that stretch from optical coherence tomography to gravitational-wave detection. In the course of conducting a series of long-duration optical-imaging experiments in a typical photonics laboratory environment that made use of a panoply of different optical sources, it became clear that at extremely low frequencies the strength of the photon-count fluctuations increased markedly with decreasing frequency. We set out to study these fluctuations in collections of experiments that were continuously conducted in a dedicated laboratory over a period of 18~months. Individual experiments with typical durations of 11.6~days were carried out under a broad variety of ambient laboratory conditions. In all cases, the photon-count fluctuations exhibited inverse-square spectral behavior that reached down to $1\:\mu$Hz, the minimum observable frequency given the limited durations of the individual experiments. 

The five classes of light sources used in our experiments included three semiconductor-based light sources: light-emitting diodes (LEDs), superluminescent diodes (SLEDs), and laser diodes (LDs). These devices emit different forms of electron--hole recombination radiation, viz., spontaneous emission, amplified spontaneous emission (ASE), and stimulated emission, respectively \citep[see, for example, Ref.][Secs.~18.1, 18.2E, and 18.4]{saleh19}. We also examined the light emitted from incandescent sources, which rely on a heated filament that emits blackbody radiation \citep[Sec.~14.4B]{saleh19}. Finally, we studied a source that operates on the basis of phosphor luminescence initiated by beta rays emitted from tritium ($^3$H) atoms \citep[][Sec.~14.5A]{saleh19}. Two types of photodetection systems were employed: a single-photon avalanche diode (SPAD) module and a photomultiplier-tube (PMT) system \citep[see Ref.][Secs.~19.4C and 19.1A, respectively]{saleh19}. 

The emphasis of our study is on the second-order statistics of these long-term photon-count fluctuations. As such, our attention is directed principally to photon-count $z$-score sample functions and their associated photon-count periodograms and scalograms. 

\section{Experiments} \label{Exp}
In this section, we report the laboratory configuration and specify the single- and dual-detector configurations used in our experiments. We also provide details about our optical sources, photodetection systems, source temperature monitoring and control systems, and data analysis and computational procedures.

\subsection{Laboratory environment} \label{ssec:labconfig}
The experiments were conducted in a standard-issue laboratory space in the Photonics Center at Boston University.  The laboratory was a closed-plan, 75-m$^2$ facility that contained an optical bench located near its center, on which all experiments were carried out. Standard commercial photonic and electronic instrumentation, including light sources, photodetectors, power supplies, and computers, were placed on the optical bench or on shelving above it, or on a separate bench to its side.  Equipment pilot lights were  extinguished or covered.  

The laboratory space was dedicated exclusively to this project during its 18-month duration. Each individual experiment took place over an 11.6-day, uninterrupted period, during which janitorial services were suspended and all personnel (including the experimenters) were excluded from the laboratory. Experiments were conducted when the HVAC (``bang-bang'' heating, ventilation, and air conditioning) system was engaged to provide the laboratory with heating or cooling, as well as when the HVAC system was disabled. Analysis revealed that the photon-counting (and temperature) fluctuations observed under all three conditions (HVAC in heating mode, HVAC in cooling mode, and HVAC disabled) were indistinguishable. Similarly, the fluctuations exhibited no discernible dependence on the season when the data were collected (fall, winter, spring, summer). The photon-counting measurements were conducted concomitantly with a series of temperature experiments, the details of which are reported separately \citep{lowen2019ambient}.

\subsection{Single- and dual-detector configurations} \label{Exp:1+2}
\begin{figure}[htb!]
\centering \includegraphics[width=\linewidth]{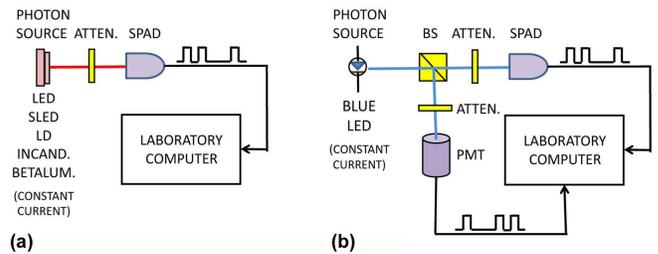}
\caption{Experimental arrangements. (a) Five classes of photon sources were used in the experiments reported here: LEDs (red, near-infrared, and blue), SLEDs (red and near-infrared), LDs (red and near-infrared), incandescent sources (white), and a betaluminescent source (green). All sources were operated in constant-current mode. 
The light from the source was attenuated and directed to a SPAD module. The source was placed at an arbitrary distance from the module and the attenuator was  adjusted such that the detected photon flux was in the vicinity of 200\,000~s$^{-1}$. The SPAD module served to detect individual photons and to convert them into idealized electronic pulses (as schematized in the figure), which were then sent to a laboratory computer for processing.
(b) In several auxiliary experiments, light from a constant-current-driven blue LED was split into two beams at a cube  beamsplitter (BS). Following attenuation that reduced the detected photon flux in each beam to roughly 200\,000~s$^{-1}$, one of the beams was directed to a SPAD module while the other impinged on a photomultiplier-tube (PMT) system. The photon-count registrations from both detectors were sent to a laboratory computer for processing.} \label{setup}
\end{figure}

The experimental arrangements used in our studies are schematically illustrated in Fig.~\ref{setup}. Most of the results reported in this paper made use of the configuration displayed in Fig.~\ref{setup}(a). The sources used in those experiments included LEDs, SLEDs, LDs, and incandescent sources, all operated in constant-current mode, as well as a betaluminescent source.
The LEDs, incandescent sources, and betaluminescent source were operated at ambient temperature whereas the SLEDs and LDs were operated at reduced temperatures and stabilized by a thermoelectric cooler. After passing through an attenuator, the light from the source was incident on a SPAD module that detected individual photons and converted them into idealized electronic pulses, as schematically shown in  Fig.~\ref{setup}(a). The nature of the results was independent of the photon flux; for convenience, the attenuator was adjusted such that the detected photon flux was about 200\,000~s$^{-1}$ for all sources.

A number of auxiliary experiments were carried out using the configuration portrayed in Fig.~\ref{setup}(b). The object of those experiments was to affirm that the results observed in the principal series of experiments did not arise from the behavior of the SPAD itself. In the auxiliary experiments, light from a blue LED driven by a constant current source was split into a pair of beams at a cube beamsplitter (Thorlabs BS017). 
One of the beams impinged on the SPAD module while the other was directed to a PMT system. Fixed and variable attenuators (Thorlabs NE40A and NDC-50C-4M, respectively) reduced the detected mean photon rates to approximately 200\,000~s$^{-1}$ in each beam. 

The experimental equipment and approach used in the studies presented here are similar to those employed by Mohan \citep{mohan10}, who made use of single- and dual-SPAD configurations similar to those portrayed in Fig.~\ref{setup} to study the behavior of the light emitted by LEDs, SLEDs, and LDs under constant-current conditions. He also studied the behavior of the light emitted by LEDs and incandescent sources under constant-voltage conditions. Mohan made use of $z$-score sample functions and periodograms to examine the concomitantly recorded photon counts, device current, device voltage, and device temperature, with photon flux as a parameter. He also investigated the correlation coefficients among the various measured quantities.

The principal distinction between our experiments and those of Mohan is the duration of the individual experiments. We extended these to 11.6~days, which enabled us to observe fluctuations at frequencies as low as 1~$\mu$Hz and also allowed us to increase the statistical accuracy of our results.
Our long-duration studies were motivated by Mohan's findings that the photon-count low-frequency spectrum revealed hints of $1/f^\alpha$ behavior. We set out to investigate this observation and to determine the value of $\alpha$, a task that required experiments of longer duration.  Less significant ways in which our experiments differed from those of Mohan are: (1) we used two different photon-detection systems (a SPAD module and a PMT system), enabling the detected photon counts to be simultaneously recorded and compared; (2) we made use of two independent thermometry systems (Thorlabs and ThermoWorks) so that temperature measurements could be simultaneously recorded and compared; and (3) we used an integration time of 5~s rather than 1~s. The sources we investigated were all operated under constant-current conditions, a configuration that was determined to be more stable than operation under constant-voltage conditions.

\subsection{Optical sources} 
The light sources used in our experiments, all of which were commercially available, are cataloged below. The various sources made use of a broad range of operating principles and had different  optical spectra, apertures, radiation patterns, and coherence properties.

\paragraph*{LEDs.} Experiments were conducted using the following devices: an AlInGaP red LED with a center wavelength of 630~nm (Kingbright WP1513SURC/E), an AlGaAs near-infrared LED with a center wavelength of 850~nm (Kingbright WP7113SF7C), and an InGaN blue LED with a center wavelength of 465~nm (Kingbright WP7113PBC/Z).
Electrical power to the the LEDs was provided by a driver (Keithley Instruments Sourcemeter 2400) that could be switched between constant-current and constant-voltage operation. A digital voltmeter/ammeter (Hewlett--Packard 34401A multimeter) with an internal integration time of $\frac16$~s served to measure the current through, and voltage across, the devices; the measured current and voltage were fed to a laboratory computer via a National Instruments General Purpose Interface Bus (GPIB), at a sampling interval of 1~s. 

\paragraph*{SLEDs.}  Experiments were conducted using the following devices: a single-transverse-mode red SLED with a center wavelength of 677.1~nm and a FWHM of 7.7~nm (Superlum SLD-260-MP1-TO9-PD), and a single-transverse-mode near-infrared SLED with a center wavelength of 839.9~nm and a FWHM of 22.8~nm (Superlum SLD-380-MP-TO9-PD). 
The SLEDs were operated under constant-current conditions and were powered by the constant-current source built into the Laser Diode and Temperature Controller (Thorlabs ITC502).
 
\paragraph*{LDs.}  Experiments were conducted using the following devices: a multiquantum-well, single-longitudinal mode AlInGaP red LD with a center wavelength of 635~nm (Hitachi HL6312G/13G); and a multiquantum-well, single-longitudinal-mode AlGaAs near-infrared LD with a center wavelength of 830~nm (Hitachi HL8325G). As with the SLEDs, the LDs were operated  under constant-current conditions and were powered by the constant-current source built into the Laser Diode and Temperature Controller (Thorlabs ITC502).
 
\paragraph*{Incandescent lamps.}  Experiments were conducted using the following lamps:  a grain-of-wheat lamp operated at a current of $\approx 50$~mA; and a 25-W, clear-bulb, decorative incandescent lamp (General Electric product code 12983) operated well below its specified operating power. The same driver and digital voltmeter/ammeter used to power the LEDs were used to power the incandescent lamps, both of which emitted white light.

\paragraph*{Betaluminescent source.} Experiments were conducted using a thin, tritium-gas-filled glass vial whose inner surface was coated with a green luminescent phosphor (Truglo TG19G). This source does not make use of an  external source of electrical power. 

\subsection{Photodetection systems} 
The following photodetection systems were used in our experiments: 

\paragraph*{SPAD module.}
The SPAD module (Perkin--Elmer SPCM-ARQ-15-FC) comprises a silicon single-photon avalanche diode, a discriminator, and a pulse shaper. The active area of the device is a circle of roughly 180-$\mu$m diameter with a spectral response in the wavelength range 400--1060~nm. A peak photon-detection efficiency of $>65\%$ is achieved at $650$ nm. The electrical power to the SPAD module was provided by a dc power supply that delivered 5V (Hewlett-Packard E3631A). The idealized output pulses from the module, generated upon the detection of individual photons, were read into a laboratory computer using a counter/timer (National Instruments NI 6602). The numbers of pulses arriving in successive counting windows of duration $T=5$~s were recorded. The module  operated at a count rate of $\approx 200\,000$~s$^{-1}$, which is more than an order of magnitude below its saturation value. The photodetector dead time of 50~ns was negligible at this count rate, as was the afterpulse probability, which was $<10^{-4}$ at 300~ns. The photon-detection-efficiency temperature coefficient of the SPAD lies between $\pm 0.1$ and $\pm 0.3 \%/{^\circ}$C over the range 5--40${^\circ}$C. Our measurements revealed that the spectrum of the SPAD-module dark-count fluctuations behaved as $\approx 1/f^{3/2}$, which accords with the behavior observed by McDowell, Ren \& Yang \citep{mcdowell08}.

\paragraph*{PMT system.}
Experiments were also carried out using an independent head-on photomultiplier-tube (PMT) system (Hamamatsu R464) in addition to the SPAD photon-counting module. This PMT has a bialkali photocathode ($5 \: \times \: 8$ mm) and a spectral response that lies in the wavelength range 300--650 nm, with a peak response at 420 nm. It contains 12 dynode stages in a box-and-grid configuration, a 21-pin glass base, and has a gain of $6 \times 10^6$. The PMT, inserted in its socket (Hamamatsu E2762-500), was operated at an anode-to-cathode voltage of 1200~V, supplied by a high-voltage power supply (Ortec~456). The PMT output was fed to the analog input of a gated photon counter (Stanford Research Systems SR400), where it was amplified and fed to a discriminator. Following a measurement of the pulse-height distribution of the PMT used in the experiments, output pulses $< - 40$~mV were deemed to represent single-photon detections. The output of the photon counter was fed to an inverting transformer (Ortec~IT100) that was followed by a fast preamplifier (Stanford Research Systems~SR445). As with the SPAD photon-counting module, the output pulses from the fast preamplifier were read into a laboratory computer using a counter/timer (National Instruments NI~6602). The numbers of pulses arriving in successive counting windows of duration $T=5$~s were recorded. A typical bialkali-photocathode PMT sensitive in the wavelength range where our blue LED emits ($\approx 458 \pm 22$ nm) has an anode-sensitivity temperature coefficient that lies between $-0.2$ and $-0.4 \%/{^\circ}$C \citep{hakamata06,hamamatsu10}.

\subsection{Source temperature monitoring and control systems}
The temperatures and temperature fluctuations of the LEDs, incandescent sources, and betaluminescent source, all of which were operated at ambient temperature, were monitored throughout the course of each experiment with a Thorlabs (Newton, NJ) Integrated-Circuit Thin-Film-Resistor Temperature Transducer (Model AD590). This system, which provides an output current proportional to absolute temperature, has a temperature resolution of $0.01^\circ$C, repeatability of $\pm0.1^\circ$C, long-term drift of $\pm0.1^\circ$C, and an accuracy of $\pm3.0^\circ$C. Its operating range stretches from $-45^\circ$C to $+145^\circ$C. Temperature readings were recorded every $5$~s and transferred to a laboratory computer via a National Instruments GPIB (General Purpose Interface Bus). For the SLEDs and LDs, the Thorlabs temperature transducer was used in conjunction with a Thorlabs Laser Diode and Temperature Controller (Model ITC502), which incorporates a thermoelectric cooler to reduce and stabilize the source temperature via a feedback loop.  

\subsection{Data analysis and computational procedures}\label{Exp:data}
The second-order statistics of the long-term photon-count fluctuations are highlighted by examining their $z$-score time traces along with the associated periodograms and scalograms. 

\paragraph*{Photon-count $\bm z$-score time traces.} 
The $z$ score is a measure of the deviation of a random variable from the mean, specified in units of standard deviation. As examples, a $z$ score of 0 indicates that the value of the random variable is the same as that of the mean while a $z$ score of $-2$ indicates that the value of the random variable lies two standard deviations below the mean.

\paragraph*{Photon-count periodograms.} Periodograms represent estimates of the power spectral density versus frequency \citep{oppenheim09}. Photon-count periodograms for the data sets collected in our laboratory using the experimental procedures described above are reported in Sec.~\ref{Results:PCPG}. The data were not truncated in any way before calculating the periodograms to avoid introducing bias. The periodograms were computed from the raw data via a procedure similar to that described in Ref.~\citep[Sec.~12.3.9]{lowen05}, a summary of which is provided below. The data used to construct the periodograms were deliberately not detrended to avoid inadvertently diluting or eliminating the fluctuations corresponding to the power-law behavior while attempting to remove nonstationarities, as explained in Ref.~\citep[][Sec.~2.8.4]{lowen05}.

\begin{enumerate}
  \item The photon-count data were normalized by setting the mean to zero and the variance to unity.
  \item The data were multiplied by a Hann (raised-cosine) window so the values at the edges were reduced to zero; this increased the frequency exponent that could be accommodated from 2 to 6, thereby comfortably including 2, as explained in Ref.~\citep[][Secs.~12.3.9 and A.8.1]{lowen05}.
  \item The data were zero-padded to the next largest power of two.
  \item The fast-Fourier transform (FFT) was computed.
  \item The square of the absolute magnitude was calculated.
  \item The resulting values were divided by the number of points in the original data set.
  \item Values with abscissas within a factor of 1.02 were averaged to yield a single periodogram abscissa and ordinate.
\end{enumerate}

\paragraph*{Photon-count scalograms.} Daubechies 4-tap (D4T) photon-count scalograms versus scale for the same data sets are reported in Sec.~\ref{Results:PCSG}. The scalograms were computed from the raw data via a procedure similar to that described in \citep[Sec.~5.4.4]{lowen05}, a summary of which is provided below. The data were not truncated in any way and neither detrending nor normalization was used in constructing the scalograms.
\begin{enumerate}
    \item A Daubechies 4-tap wavelet transform \citep{daubechies92} was carried out, using all available scales. 
    \item All elements of the wavelet transform that spanned the edges of the data set were discarded.
    \item The square of all remaining elements were calculated; since wavelet transforms are zero-mean by construction, this allowed the wavelet variance to be directly determined.
    \item All such elements with the same scale were averaged.
\end{enumerate} 

\paragraph*{Other measures.} Short-term marginal statistics such as photon-count histograms were used to provide further confirmation that our observations were sound; correlation coefficients for all pairs of variables of interest (photon count, current, and voltage) served to elucidate the relationships among these quantities \citep{mohan10}.

\subsection{Notation}\label{Exp:notation}
\begin{enumerate}
  \item The photon-count periodogram is denoted $S_P(f)$, where $f$ is the periodogram harmonic frequency. The source temperature periodogram is denoted $S_\mathcal{T}(f)$, where $\mathcal{T}$ is the source temperature.
  \item The inverse of the periodogram frequency $f$ is denoted $T_f$. Diurnal photon-count or temperature variations, such as those that might arise from daylight leaking into the laboratory or day/night temperature differences, respectively, correspond to $T_f = 1$~d = $86\,400$~s and therefore appear in the periodogram at $f = 1/T_f = 1.16 \times 10^{-5}$~Hz and harmonics thereof.
  \item Sampling that takes place at intervals of $T_s$ results in a maximum allowed frequency $\frac12 /T_s\,$, as established by the Nyquist limit \citep{oppenheim09}. For photon-counting experiments carried out using $T_s = 5$~s, the highest allowed frequency is therefore $\frac12 / (5~\mathrm{s}) = 0.10$~Hz. 
  \item The photon-count Daubechies 4-tap wavelet variance is denoted $A_P(T)$, where $T$ is the wavelet scale.
\end{enumerate}

\section{Results} \label{Results}
The principal results pertaining to our studies of the second-order statistics of long-term photon-count fluctuations are presented in  Secs.~\ref{Results:PCPG} and \ref{Results:PCSG}, which consider photon-count periodograms and photon-count scalograms, respectively. Sections~\ref{Results:DDPG} and \ref{Results:STPG}  pertain to dual-detector periodograms and source temperature periodograms, respectively, and are designed to demonstrate that the  observed low-frequency fluctuations do not arise from noise or temperature fluctuations at the detector, nor from temperature fluctuations at the source. The results presented in this paper are drawn from 23 data sets whose durations are sufficiently long that they can access $\mu$Hz frequencies with high statistical accuracy. These 23 data sets are in turn drawn from a total of 34 data sets collected over the course of our experiments.

\begin{figure*}[htbp!]
  \centering
  \includegraphics[width=17cm]{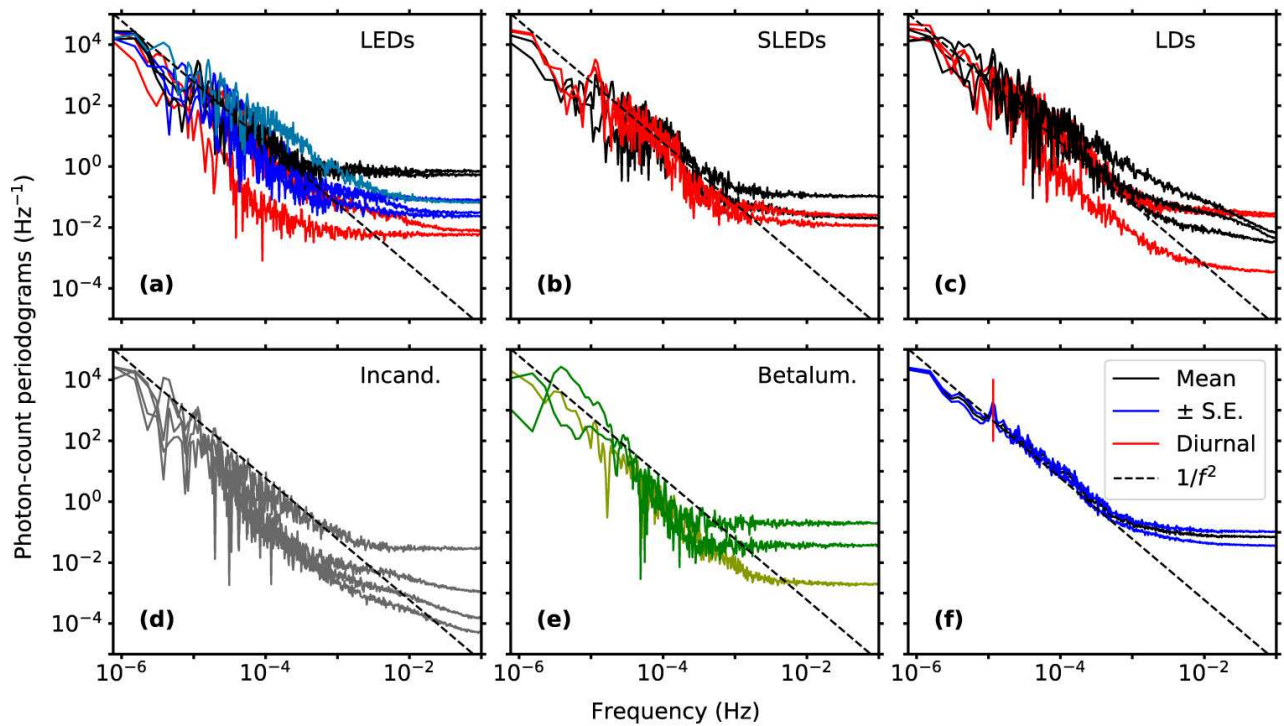}
  \caption{Photon-count periodograms plotted on doubly logarithmic coordinates. Individual normalized periodograms, $S_P(f)$ versus frequency $f$, are plotted one atop the other for five classes of optical sources. The colors in which the curves are plotted correspond to the colors of the light emitted by the individual devices (black and gray represent near-infrared and white light, respectively). The data  presented were collected using the SPAD module, except where indicated. (a)~LEDs (eight data sets) --- the cyan curve represents data collected with the PMT photodetection system, (b)~SLEDs (four data sets), (c)~LDs (six data sets), (d)~incandescent sources (four data sets), and (e)~a betaluminescent source  (three data sets) --- the yellow-green curve represents data collected with the PMT photodetection system. Twenty-three of the 25 device periodograms displayed in panels~(a)--(e) were derived from data sets of the same length (11.6~d); the other two curves, portrayed as green in panel~(e), had durations of 8.7~d and 9.3~d. (f)~Despite the fact that they were measured at different times and under different laboratory conditions, and that they correspond to five different classes of sources, the 23 periodograms follow each other sufficiently closely that they can be averaged to improve the statistical accuracy of the data. The black curve represents the averaged periodogram $\overline{S}_P(f)$ versus frequency $f$ and the two blue curves hugging it represent standard-error values ($\pm$~S.E.). The mean periodogram decreases approximately as $1/f^2$ (dashed line) over the frequency range $1.0 \times 10^{-6} \le f \le 5.0 \times 10^{-4}$~Hz.
 }
\label{bothphotonpg}
  \end{figure*}
  
\subsection{Photon-count periodograms} \label{Results:PCPG}
Photon-count periodograms for all five classes of optical sources are presented in Fig.~\ref{bothphotonpg}. 
The data were collected using the experimental configuration portrayed in Fig.~\ref{setup}(a); the periodograms were computed using the procedure described in Sec.~\ref{Exp:data}. Figures~\ref{bothphotonpg}(a)--\ref{bothphotonpg}(e) represent collections of  individual periodograms for LEDs, SLEDs, LDs, incandescent sources, and a betaluminescent source, respectively.

Averaging the 23 periodograms displayed in Figs.~\ref{bothphotonpg}(a)--\ref{bothphotonpg}(e) that are constructed from data sets of sufficient length yields the black curve shown in Fig.~\ref{bothphotonpg}(f), which exhibits improved statistical accuracy by virtue of the averaging process. To construct the mean periodogram, a threshold of 198\,048 measurements was determined to be the best compromise between including as many data sets, and as much of each, as possible; all 23 data sets were truncated to that number of measurements. At 5~s per measurement, that corresponds to $L = 990\,240$~s. The next largest power of 2 greater than 198\,048 is 262\,144, or 1\,310\,720~s. The lowest observable frequency is very close to 1~$\mu$Hz; it  could be reduced by zero-padding the data set. The highest observable frequency is the Nyquist limit $\frac12 / (5~\mathrm{s}) = 0.10$~Hz, as discussed in Sec.~\ref{Exp:notation}. However, because of averaging of the values whose abscissas lie within a factor of 1.02, the highest displayed frequency  is a bit smaller than 0.10~Hz, namely 0.098~Hz. 

The outcome of the averaging process displayed in Fig.~\ref{bothphotonpg}(f) exhibits approximately $2\frac12$ log-units of $1/f^2$ behavior (dashed line) over the frequency range $1.0 \times 10^{-6} \le f \le 5.0 \times 10^{-4}$~Hz, corresponding to 33~min $\le T_f \le 11.6$~d. The small hump in the curve at $1.16 \times 10^{-5}$~Hz, designated by a vertical red line segment, corresponds to diurnal variations (1~d = 86\,400~s). Since the periodogram displayed in Fig.~\ref{bothphotonpg}(f) flattens out at $f \gtrsim 5.0 \times 10^{-4}$~Hz, corresponding to $T_f \lesssim 33$~min, sampling the photon count at intervals shorter than $T_s = \frac12 T_f \approx 16$~min does not provide additional information.

The periodograms displayed in Figs.~\ref{bothphotonpg}(a)--\ref{bothphotonpg}(e) reveal that, for each type of source, the low-frequency spectra for all of the individual devices exhibit $1/f^2$ behavior, whatever their optical spectra. Moreover, the averaged periodogram shown in Fig.~\ref{bothphotonpg}(f) reveals that the low-frequency spectra for all of the different types of sources behave similarly, whatever their light-generation mechanism, radiation pattern, and coherence properties. These observations are confirmed by the photon-number scalograms that will be presented in Fig.~\ref{bothphotonwv}. 

\begin{figure*}[htbp!]
\centering
\includegraphics[width=15.5cm]{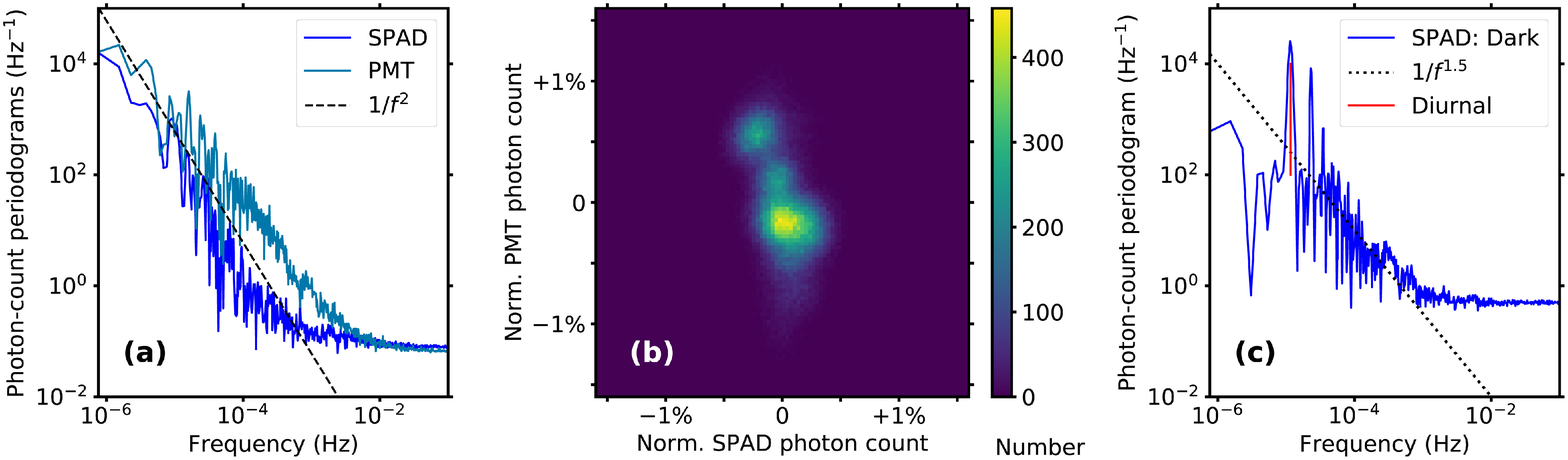}
\caption{Photon-count measurements obtained from an experiment using two independent photodetection systems, configured as in Fig.~\ref{setup}(b).  (a)~Normalized photon-count periodograms, $S_P(f)$ versus frequency $f$, obtained using the SPAD module and the PMT photodetection system, plotted on doubly logarithmic coordinates. These curves, which were among the eight LED curves presented in Fig.~\ref{bothphotonpg}(a), are drawn from an experiment of duration 11.6~d (file names 20111214a and 20111214b).
Both curves exhibit approximate $1/f^2$ spectral behavior (dashed line); close examination reveals that the details of their fluctuations are substantially similar. However, the SPAD module exhibits greater baseline noise than the PMT system. (b)~Joint photon-count histogram displaying the photon counts registered by the SPAD (abscissa) and the PMT (ordinate),  normalized to their respective values of the mean count, in a sequence of consecutive counting windows of duration $T = 5$~s. The joint histogram, constructed from a 7.51-d subset of the 11.6-d data set (file names 20111214a and 20111214b), exhibits positive overall correlation and a narrow dominant peak, indicating that the two detectors register similar results.
(c)~Normalized photon-count periodogram for the SPAD module in the absence of deliberate illumination, plotted on doubly logarithmic coordinates.
This curve is drawn from an experiment of duration 11.6~d (file name 20101227). Over the central frequency region, the spectrum behaves as $1/f^{3/2}$ (dotted curve), which differs from the spectra that emerge in the presence of illumination, as displayed in panel~(a) and in Fig.~\ref{bothphotonpg}. 
The peak in the curve at $1.16 \times 10^{-5}$~Hz, designated by the vertical red line segment, and other peaks at harmonics thereof, correspond to diurnal variations (1~d = 86\,400~s) resulting from unintended residual leakage of a small amount of daylight into the laboratory.
}
\label{APD+PMT}
\end{figure*}
 
\subsection{Dual-detector periodograms} \label{Results:DDPG}

We now demonstrate that two independent photon-counting systems, one based on the SPAD module used to collect the data displayed in Fig.~\ref{bothphotonpg}, and the other based on the PMT system, yield similar results even though the two photodetection systems behave very differently.

Using the experimental configuration depicted in Fig.~\ref{setup}(b), in which data is simultaneously acquired by both detectors, we obtain the normalized photon-count periodograms displayed in Fig.~\ref{APD+PMT}(a). The detailed behavior of the two curves is quite similar and both exhibit approximate inverse-square spectral behavior (dashed line). The principal distinction between the periodograms is the frequencies at which they flatten out. The baseline noise for the SPAD becomes appreciable at $f \approx 5 \times 10^{-4}$~Hz while the PMT is capable of registering fluctuations over an additional log-unit of frequency. In spite of the fact that it is widely used because of  convenience, the SPAD is substantially noisier than the PMT \citep[Sec.~19.6]{saleh19}. To provide perspective, we mention that the cutoff frequencies instigated by detector noise are far more limiting than those arising from detector response time and Nyquist sampling. 

The joint photon-count histogram for the two photodetectors, plotted on axes whose values are normalized by their respective mean photon counts, is displayed in Fig.~\ref{APD+PMT}(b). The data have positive overall correlation and exhibit a narrow dominant peak whose width lies well within 1\% of the normalized photon-count values. The numbers of observed events are indicated by the color key at the right side of Fig.~\ref{APD+PMT}(b). With a counting time $T = 5$~s, and an experiment of duration 7.51~d, the mean numbers of photon counts and their variances for the SPAD and PMT photodetection systems were, respectively, $\overline{n}_\mathrm{SPAD} = 985\,370$, $\sigma^2_\mathrm{SPAD} = 3\,256\,365$; and $\overline{n}_\mathrm{PMT} = 995\,441$,  
 $\sigma^2_\mathrm{PMT} = 15\,541\,808$. The mean photon counts differ slightly because the detectors were not presented with the same values of the photon flux and have different quantum efficiencies; hence, the plot in Fig.~\ref{APD+PMT}(b) makes use of axes normalized to their mean values. 
 
 It is apparent from Figs.~\ref{APD+PMT}(a) and \ref{APD+PMT}(b) that the results obtained using the two photodetectors are  quite similar, thereby confirming that our measurement methods are sound and that our results are consistent. Moreover, since the SPAD and PMT systems operate on the basis of different principles, and exhibit quite different noise properties, temperature dependencies, and quantum efficiencies, the results presented in Fig.~\ref{APD+PMT}(a) indicate that the $1/f^2$ behavior of the photon-count periodograms displayed in Fig.~\ref{bothphotonpg} do not have their origins in noise or temperature fluctuations at the detector. 
 
 Finally, we note that in the absence of light, the SPAD module exhibits a spectrum that behaves as $\approx 1/f^{3/2}$ at these low frequencies, as displayed in Fig.~\ref{APD+PMT}(c), rather than as $\approx 1/f^2$, as it does in the presence of light.

\subsection{Source temperature periodograms}
\label{Results:STPG}

\begin{figure}[htbp!]
\centering
\includegraphics[width=\columnwidth]{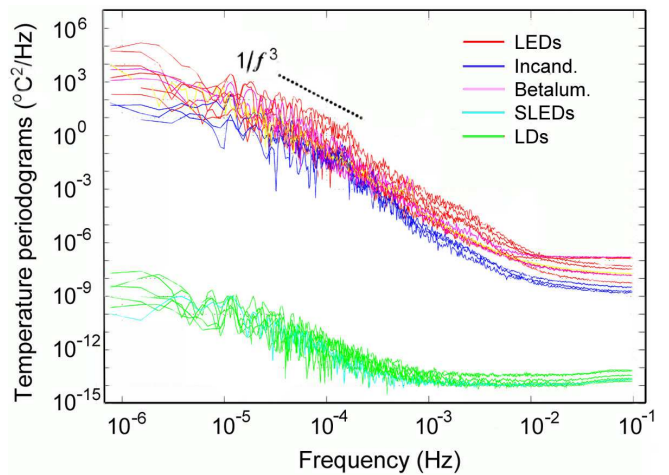}
\caption{Source temperature periodograms collected during experiments with various photon sources. The upper curves, plotted one atop the other, are unnormalized periodograms, $S_\mathcal{T}(f)$ versus $f$, recorded during individual experiments with LEDs, incandescent sources, and a betaluminescent source, which are operated at ambient temperature. The lower curves portray unnormalized periodograms recorded during individual experiments with SLEDs and LDs, for which temperature control was imposed, again plotted one atop the other. In the region where both sets of curves are decreasing, the lower cluster lies 10--12 orders of magnitude below the upper cluster, indicating that source temperature fluctuations are not responsible for the behavior of the photon-count periodograms.}
\label{temppgs}
\end{figure}

We now proceed to establish that source temperature fluctuations do not appear to be  responsible for the $1/f^2$ form of the photon-count periodograms displayed in Fig.~\ref{bothphotonpg}. In Fig.~\ref{temppgs} we present a collection of unnormalized source temperature periodograms collected with the Thorlabs temperature sensor that were concomitantly recorded with the photon-count periodograms. The upper cluster of curves are temperature periodograms for experiments carried out with LEDs, incandescent sources, and a betaluminescent source, where the source temperature was the same as the recorded ambient temperature.
The best fit to this cluster of curves over the frequency range $1.0
\times 10^{-5} \le f \le 5.0 \times 10^{-3}$~Hz is a straight line representing behavior of the form $1/f^3$, in accord with previous observations \citep{lowen2019ambient}.

The lower cluster of curves are temperature periodograms for experiments carried out with SLEDs and LDs, where temperature control of the source was instituted and the local temperature of the source was recorded. This cluster of periodograms lies 10--12 orders of magnitude below the upper cluster in the region where both sets of curves are decreasing.

Since the temperature fluctuations in the upper cluster of curves are a trillion times stronger than those associated with the lower cluster, and since the temperature periodograms in the upper cluster do not project the $1/f^2$ behavior that is the hallmark of the photon-count periodograms displayed in Fig.~\ref{bothphotonpg}, we conclude that source temperature fluctuations are not responsible for the behavior of the photon-count periodograms.

\subsection{Photon-count scalograms}
\label{Results:PCSG}
\begin{figure*}[ht]
  \centering
  \includegraphics[width=17cm]{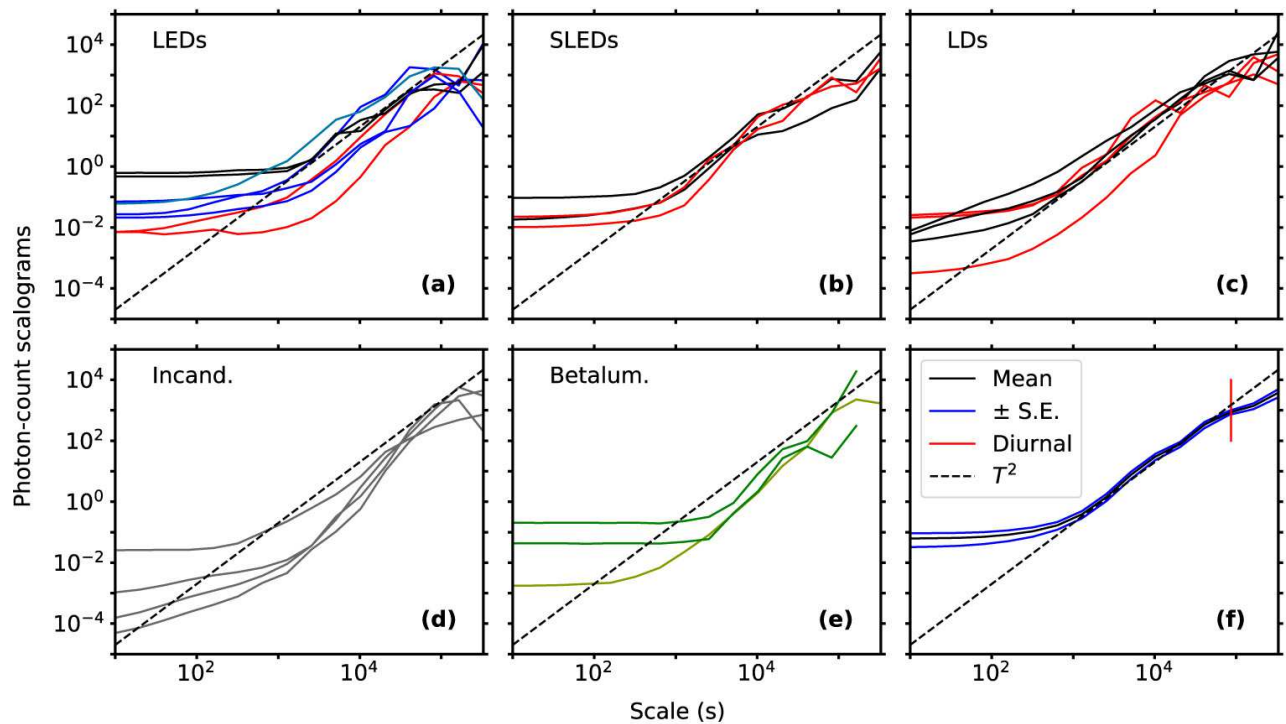}
  \caption{
  Photon-count scalograms plotted on doubly logarithmic coordinates. Individual normalized Daubechies 4-tap wavelet variance curves, $A_P(T)$ versus scale $T$, are plotted one atop the other for the five classes of optical sources. The colors in which the curves are plotted correspond to the colors of the light emitted by the individual devices (black and gray represent near-infrared and white light, respectively). The data  presented were collected using the SPAD module, except where indicated. (a)~LEDs (eight data sets) --- the cyan curve represents data collected with the PMT photodetection system, (b)~SLEDs (four data sets), (c)~LDs (six data sets), (d)~incandescent sources (four data sets), and (e)~a betaluminescent source  (three data sets) --- the yellow-green curve represents data collected with the PMT photodetection system. Twenty-three of the 25 device scalograms displayed in panels~(a)--(e) were derived from data sets of the same length (11.6~d); the other two curves, portrayed as green in panel~(e), had durations of 8.7~d and 9.3~d. (f)~Despite the fact that they were measured at different times and under different laboratory conditions, and that they correspond to five different classes of sources, the 23 scalograms follow each other sufficiently closely that they can be averaged to improve the statistical accuracy of the data. The black curve represents the averaged Daubechies 4-tap wavelet variance $\overline{A}_P(T)$ versus scale $T$ and the two blue curves hugging it represent standard-error values ($\pm$~S.E.). The mean scalogram increases approximately as $T^2$ (dashed line) over the range $1.0 \times 10^3 \le T \le 1.0 \times 10^5$~s, in correspondence
  with the $1/f^2$ decrease of the averaged periodogram portrayed in Fig.~\ref{bothphotonpg}(f).
 }
  \label{bothphotonwv}
  \end{figure*}

Photon-count scalograms for the five classes of optical sources investigated are presented in Fig.~\ref{bothphotonwv}. 
The data were collected using the experimental configuration set forth in Fig.~\ref{setup}(a) and these measures are extracted from the same raw data as those used to generate the photon-count periodograms portrayed in Fig.~\ref{bothphotonpg}.  Figures~\ref{bothphotonwv}(a)--\ref{bothphotonwv}(e) represent individual scalograms for LEDs, SLEDs, LDs, incandescent sources, and a betaluminescent source, respectively, plotted one atop the other. Figure~\ref{bothphotonwv}(f) displays the average of the 23 individual scalograms presented in Figs.~\ref{bothphotonwv}(a)--\ref{bothphotonwv}(e) that were constructed from data sets of the same length. The scalograms were computed using the procedure described in Sec.~\ref{Exp:data}. 

As discussed in Ref.~\citep[Sec.~12.3.2]{lowen05}, the normalized \emph{ordinary} variance (Fano factor), defined as the count variance-to-mean ratio $F_P(T) = \sigma^2_n/\overline{n}$, is a biased estimator that cannot increase faster than $T^1$, where $T$ is the duration of the counting window. We therefore rely instead on a version of the normalized \emph{wavelet} variance that is not subject to this limitation. In particular, we make use of the Daubechies 4-tap wavelet variance (D4TWV)  $A_P(T)$ because of its insensitivity to constant values and linear trends \citep{vetterli95,teich95}, and because of the extended power-law exponent it accommodates. As explained in Ref.~\citep[Sec.~5.2.5]{lowen05}, the observed power-law increase of the D4TWV as $A_P(T) \propto T^2$ for our scalograms lies well below growth as $T^5$, where 5 is the maximum exponent permitted for the increase of the D4TWV with scale and signals nonstationary behavior. We note that the normalized Haar-wavelet variance (NHWV), though it has excellent properties and is widely used, can rise no faster than $T^3$ and is not insensitive to linear trends \citep[Sec.~12.3.8]{lowen05}. 

As with the periodograms displayed in Figs.~\ref{bothphotonpg}(a)--\ref{bothphotonpg}(e), the scalograms shown in Figs.~\ref{bothphotonwv}(a)--\ref{bothphotonwv}(e) follow each other sufficiently closely that they may be averaged to improve  statistical accuracy. The outcome of the averaging process, displayed in Fig.~\ref{bothphotonwv}(f), 
increases approximately as $T^2$ (dashed line) over the range $1.0 \times 10^3 \le T \le 1.0 \times 10^5$~s.
The maximum permitted scale, corresponding to 1 sample, is 327\,680~s or 3.8~d. The observed square-law growth of the averaged scalogram presented in Fig.~\ref{bothphotonwv}(f) is a mirror image of the square-law decrease of the observed averaged periodogram portrayed in Fig.~\ref{bothphotonpg}(f). 

The mean wavelet variance $\overline{A}_P(T)$ shown in Fig.~\ref{bothphotonwv}(f) exhibits low- and high-scale cutoffs at $T_A \approx 1.0 \times 10^3$~s and $T^\prime_A\approx 1.0 \times 10^{5}$~s, respectively. Analogously, the mean periodogram $\overline{S}_P(f)$ portrayed in Fig.~\ref{bothphotonpg}(f) displays low- and high-frequency cutoffs at $f^\prime_S \approx 1.0 \times 10^{-6}$~Hz and $f_S  \approx 5.0 \times 10^{-4}$~Hz, respectively. Thus, the \emph{low-frequency/high-scale} cutoff product $f^\prime_S  T^\prime_A  \approx  0.1$ and the \emph{high-frequency/low-scale} cutoff product $f_S T_A  \approx  0.5$; these values are in rough agreement, as expected.   

In summary, just as with the periodograms, the scalograms displayed in Figs.~\ref{bothphotonwv}(a)--\ref{bothphotonwv}(e) reveal that all of the individual devices associated with each type of source behave similarly, whatever their optical spectra. And, just as with the averaged periodogram, the averaged scalogram shown in Fig.~\ref{bothphotonwv}(f) confirms that the scalograms for the various types of sources behave similarly, whatever their light-generation mechanism and whatever the statistical properties of the light they generate. The consistency of the scalograms and periodograms confirms that the computations for both measures are sound.
 
\section{Discussion} \label{Discussion}
A wide body of research exists detailing the coherence properties and photon-count statistics of various sources of light \citep{saleh78,perina85,teich88}. Some forms of light, such as {\v C}erenkov radiation from a random stream of charged particles, exhibit power-law spectra \citep{lowen91pra}. However, the observation of $1/f^2$  photon-count fluctuations at baseband is unexpected and novel, and also perplexing. 

\begin{figure*}[ht]
 \includegraphics[width=17cm]{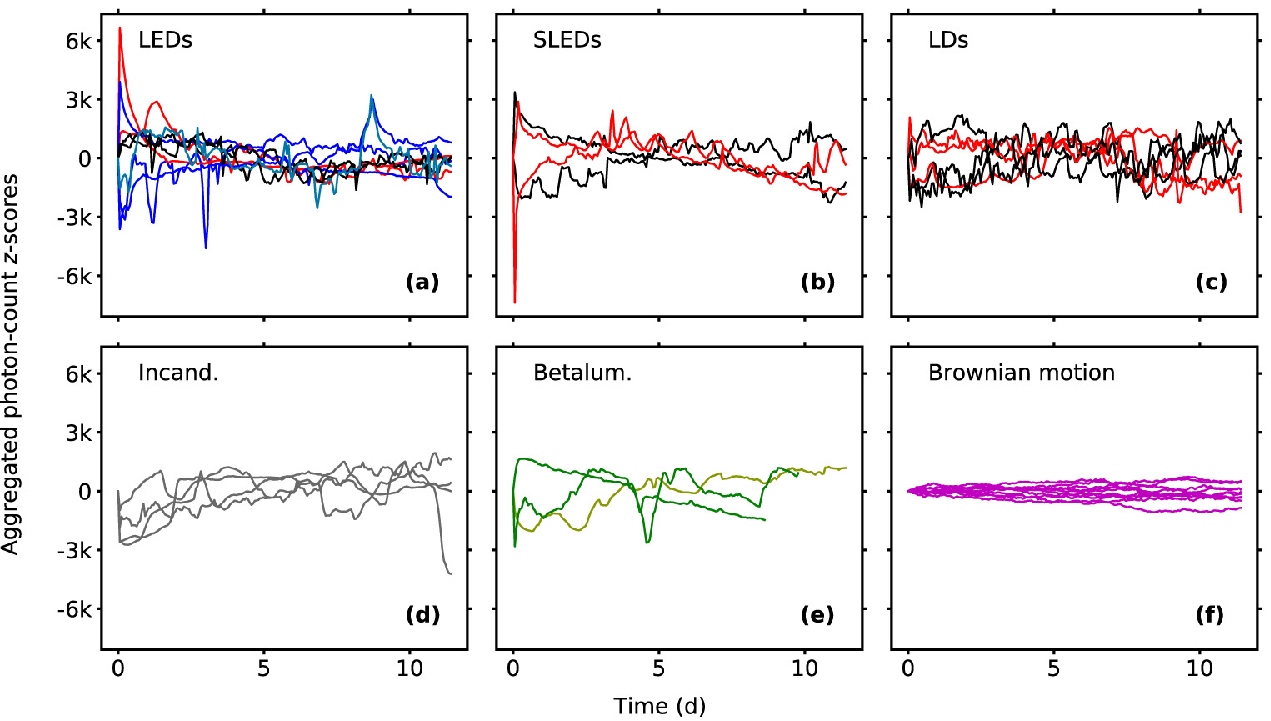}
\caption{Individual photon-count sample functions, standardized as aggregated $z$ scores, plotted one atop the other for the five classes of optical sources investigated. The initial value of each curve was set to zero to facilitate comparison. The colors in which the curves are plotted correspond to the colors of the light emitted by the individual devices (black and gray represent near-infrared and white light, respectively; magenta represents simulated Brownian motion). The data  presented were collected using the SPAD module, except where indicated. (a)~LEDs (eight data sets) --- the cyan curve represents data collected with the PMT photodetection system, (b)~SLEDs (four data sets), (c)~LDs (six data sets), (d)~incandescent sources (four data sets), and (e)~a betaluminescent source  (three data sets) --- the yellow-green curve represents data collected with the PMT photodetection system. All of the sample functions in this figure represent data sets of the same length (11.6~d) with the exception of the two green curves in panel~(e), whose durations are 8.7~d and 9.3~d. Though measured at different times and under different laboratory conditions, all of the traces exhibit irregular long-time-scale fluctuations, whatever the type of source or its spectrum. (f) Ten simulated sample functions of ordinary Brownian motion (magenta) whose  variances are commensurate with the data. The Brownian curves are considerably smoother than the photon-count fluctuations displayed in panels~(a)--(e).
}
\label{brown}
\end{figure*}

Since an inverse-square-law spectrum at low frequencies often signals the presence of an underlying source of Brownian motion 
\cite[see, for example, Ref.][Sec.~2.4.2]{lowen05}, it is useful to examine the time course of the photon count to ascertain if that is the case here.
Individual aggregated photon-count $z$ scores for the five classes of optical sources examined are displayed in Figs.~\ref{brown}(a)--\ref{brown}(e). The time traces were calculated by summing disjoint sets of 1000 consecutive data values, yielding one aggregated number corresponding to 5000~s (83.33~min) of photon arrivals for each set. For each curve, the value of the first aggregated number was subtracted from all aggregated numbers for that curve, thereby yielding a curve whose initial value is zero. For purposes of comparison, the ten magenta curves presented in Fig.~\ref{brown}(f) portray simulated sample functions for 1D ordinary Brownian motion with commensurate values of the variance. Comparison of the photon-count traces in Figs.~\ref{brown}(a)--\ref{brown}(e) with those in Fig.~\ref{brown}(f) reveals that the photon-count curves are distinctly different in character from those for ordinary Brownian motion. 
It is worth noting that the data displayed in Figs.~\ref{brown}(a), \ref{brown}(d), and \ref{brown}(e) were collected from devices operated at ambient temperature, whereas those displayed in Figs.~\ref{brown}(b) and \ref{brown}(c) were collected from cooled devices; no dramatic distinction between the curves for the two classes of data is apparent.

Close examination of the curves presented in  Figs.~\ref{brown}(a)--\ref{brown}(e) reveals that the characteristics of the aggregated photon-count $z$-scores appear to vary somewhat from one source class to another. In particular, we observe the following features:
\begin{enumerate}
    \item The photon-count time curves for all of the optical sources exhibit what appears to be drift; this is a manifestation of fluctuations at the lowest discernible frequencies. Riding atop these slow fluctuations are more rapid variations whose time structures assume somewhat different forms. 
    \item For some sources, particularly LEDs and SLEDs, the photon-count sample functions exhibit what would customarily be called spikes; however, at the time scales of these figures, the durations of these features are typically of the order of hours.
    \item For all sources, with the possible exception of the betaluminescent source, some of the time traces contain segments that resemble relaxation oscillations, but with periods that are typically of the order of a day. Optical sources do indeed often generate relaxation oscillations, but such behavior is not expected at these long time scales. 
    \item For the betaluminescent source, the time traces exhibit structure with scales that range from several hours to several days.
\end{enumerate} 
Despite their apparent diversity, these fluctuations all exhibit an inverse-square baseband spectrum and a concomitant square-law wavelet variance at large scales.  

The lion's share of the traces presented in Fig.~\ref{brown} were collected using the SPAD module but two traces were collected using the PMT photodetection system. It is interesting to observe that the aggregated $z$-score time traces simultaneously measured by the SPAD module and the PMT photodetection system via the experimental configuration depicted in Fig.~\ref{setup}(b) bear some  resemblance to each other, as can be seen by comparing the cyan trace with its companion blue trace in Fig.~\ref{brown}(a). This bolsters the evidence proffered in Fig.~\ref{APD+PMT}(a) that the long-time-scale fluctuations in the photon count likely do not originate at the photodetector. It is also noteworthy that the aggregated $z$ score for the SPAD dark counts (not shown) fluctuates far less than those for the photon counts in the presence of light, despite the fact that a small amount of unintended residual daylight leaks into the laboratory, as is evidenced by the diurnal fluctuations present in the periodogram displayed in Fig.~\ref{APD+PMT}(c).

The photon-count fluctuations of the betaluminescence displayed in Fig.~\ref{brown}(e) are of particular interest. It is apparent from examining the various panels in Fig.~\ref{brown} that these fluctuations do not differ fundamentally in character from those associated with more conventional sources, such as LEDs and incandescent lamps. However, the light emitted by this $^3$H source is  independent of temperature and is also  generated by a constant flux of $\beta$-decay (half-life $t_{1/2} \approx 12.32$~y). This provides further evidence that source  temperature fluctuations, and any putative fluctuations imparted by an external electrical power source, are  not responsible for the long-time-scale fluctuations in the photon count.  

Finally, from a practical perspective it is worth mentioning that photonic applications requiring a stable photon flux over the long term can make use of feedback control to modulate either the device power or a variable-transmittance optical filter that follows the source.

\section{Conclusion} \label{Conclusion}
 We have presented an assemblage of photon-count periodograms, collected under a broad variety of laboratory conditions, that display unexpected low-frequency fluctuations. The experiments were conducted using five classes of optical sources that have markedly different optical spectra, radiation patterns, and coherence properties, and that generate light via distinct mechanisms: spontaneous emission (LEDs), amplified spontaneous emission (SLEDs), stimulated emission (LDs), blackbody radiation (incandescent sources), and luminescence radiation (betaluminescent source).

 Every device we examined, in all of the classes of optical sources we investigated, yielded photon-count periodograms with a $1/f^2$ low-frequency spectral signature that extended to $< 1\;\mu$Hz. Daubechies 4-tap wavelet scalograms revealed $T^2$ scale signatures at large values of $T$ for all of the devices, in correspondence with the $1/f^2$ spectral signatures of the periodograms at small values of $f$, confirming that our computational methods are sound.
 
 Dual photon-count periodograms and aggregated $z$-score traces collected with photodetection systems that operate on the basis of different physical principles support the conclusion that the $1/f^2$ photon-count fluctuations do not arise from noise, temperature fluctuations, or power-supply fluctuations at the photodetector, and their consistency verifies that our measurement methods are sound. The features of the measured source temperature periodograms, together with the long-time-scale photon-count fluctuations present in betaluminescence, establish that the photon-count fluctuations are not instigated by fluctuations in the temperature of the source or in its external power supply.
 
The aggregated photon-count $z$-score time traces exhibit considerable variability across sources and do not resemble those of ordinary Brownian motion. It is important to definitively establish the origins and nature of these photon-count fluctuations. The purpose of the present paper is more modest, however: it is to report the existence of these slow photon-count fluctuations and to provide evidence that their inverse-square spectral behavior at baseband is conserved across a broad variety of optical sources.
 
\begin{acknowledgments}
We thank Jeffrey~H. Shapiro for valuable discussions. We are grateful to the Boston University Photonics Center and its director, Thomas Bifano, for financial and logistical support. The Photonics Center provided a dedicated laboratory in which we could control, restrict, and suspend climate control (heating and air conditioning), janitorial services, and the access of personnel. Individual experimental runs, which typically lasted approximately two weeks, were conducted continuously over a period of 18 months, from May 2010 until December 2011. 
\end{acknowledgments}

\bibliography{main}

\end{document}